\documentclass{emulateapj}

\newcommand{\tauff}{\tau_{ff}}
\newcommand{\msun}{M_\odot}
\newcommand{\etal}{et~al.}
\newcommand{\be}{\begin{equation}}
\newcommand{\en}{\end{equation}}

\begin{document}

\title{Rapid Molecular Cloud and Star Formation: Mechanisms \& Movies}

\author{Fabian Heitsch\altaffilmark{1}}
\author{Lee Hartmann\altaffilmark{1}}
\altaffiltext{1}{Dept. of Astronomy, University of Michigan, 500 Church St., 
                 Ann Arbor, MI 48109-1042, U.S.A}
\lefthead{Heitsch \& Hartmann}
\righthead{Rapid Cloud and Star Formation}

\begin{abstract} 
We demonstrate that the observationally inferred rapid onset of star formation after 
parental molecular clouds have assembled can be achieved by flow-driven cloud formation
of atomic gas, using our previous three-dimensional numerical simulations.
We post-process these simulations to approximate CO formation, which allows us to
investigate the times at which CO becomes abundant
relative to the onset of cloud collapse. We find that global gravity in a finite cloud
has two crucial effects on cloud evolution.  (a) Lateral collapse
(perpendicular to the flows sweeping up the cloud) leads to rapidly increasing column densities
above the accumulation from the one-dimensional flow.  This in turn allows
fast formation of CO, allowing the molecular cloud to ``appear'' rapidly. 
(b) Global gravity
is required to drive the dense gas to the high pressures necessary to form  
solar-mass cores, in support of 
recent analytical models of cloud fragmentation. While the clouds still appear 
``supersonically turbulent'', this turbulence is relegated to playing a secondary role,
in that it is to some extent a consequence of gravitational forces. 
\end{abstract}

%%%%%%%%%%%%%%%%%%%%%%%%%%%%%%%%%%%%%%%%%%%%%%%%%%
%
%\section{Introduction}
%
%%%%%%%%%%%%%%%%%%%%%%%%%%%%%%%%%%%%%%%%%%%%%%%%%%

\section{Introduction}\label{s:intro}

One motivation for pursuing
a picture of molecular cloud formation by large-scale flows was 
to resolve the ``crossing time problem'', i.e. the fact that the age spreads
of young stars are often much shorter than lateral dynamical timescales
(\citealp{1999ApJ...527..285B}; \citealp{2001ApJ...562..852H}, HBB01).  
In this picture, dense
star-forming clouds are produced as flows -- expanding H{\small II} regions,
stellar wind bubbles, supernova explosions, spiral density waves (e.g. \citealp{2006MNRAS.365...37B}) 
or global
gravitational instabilities (e.g. \citealp{2007ApJ...671..374Y}) -- sweep up 
and condense diffuse interstellar gas.  The crossing time problem is eliminated
because information is not being transmitted along the long dimension of the cloud.

A major attraction of the notion of flow-driven molecular cloud formation 
is that we see several examples of it in the solar neighborhood, such as
in Cep OB2 \citep{1998ApJ...507..241P}, in addition to the well-known cloud and star
formation in galactic spiral arms due to gas inflow 
(e.g. \citealp{1979ApJ...231..372E,2007ApJ...668.1064E}; see e.g. 
\citealp{2003ApJ...599.1157K} and \citealp{2007MNRAS.376.1747D} for numerical evidence).
While it is not always easy to identify specific driving sources
in all cases, this is not surprising given
the complexity expected as a result of the interaction of neighboring flows.
In addition, the presence of molecular clouds well out of the galactic plane 
(e.g., Orion) clearly suggests the need for some kind of driving.
Given the extensive impact that massive stars have on the interstellar medium --
H{\small II} regions, stellar wind impacts, and ultimately supernova explosions -- it is difficult
to see how further creation of new star-forming locales by flows with scales of several to
tens of pc (or even kpc, in the case of spiral arms) could be avoided.

Star-forming clouds in this picture become somewhat accidental associations of gas
which are not in virial equilibrium, although rough energy equipartition is expected 
(HBB01), in adequate agreement with observations \citep{2006MNRAS.372..443B}.
Indeed, it is very difficult to prevent global gravitational collapse motions 
in finite clouds of many Jeans masses \citep{2004ApJ...616..288B}, consistent
with the empirical evidence for rapid star formation following cloud formation
(HBB01).  

While global gravitational contraction provides a plausible mechanism for assembling
protocluster gas and stars and even overall cloud morphology \citep{2007ApJ...654..988H}, 
it poses difficulties for local collapse; non-linear perturbations are probably required
to avoid sweep-up in overall collapse modes \citep{2004ApJ...616..288B}.
The necessity of producing star-forming clumps through turbulent motions has long
been recognized \citep{1981MNRAS.194..809L}, based on a large number of numerical simulations 
over the years (see review by \citealp{2004RvMP...76..125M}). However, the source
of this turbulence has been a matter of debate.  The flow-driven cloud formation picture
may provide an answer, in that the dynamical instabilities coupled with rapid cooling and
thermal instability which naturally result at the shock interface between driving material
and swept-up gas generate turbulence and non-linear density fluctuations 
(\citealp{2002ApJ...564L..97K}; \citealp{2002Ap&SS.281...67I}; \citealp{2005A&A...433....1A}; 
\citealp{2005ApJ...633L.113H}; \citealp{2006ApJ...643..245V}; \citealp{2006ApJ...648.1052H}; 
\citealp{2007ApJ...657..870V}; \citealp{2007A&A...465..445H}; \citealp{2007A&A...465..431H}; 
\citealp{2008ApJ...674..316H}; \citealp{2008A&A...486L..43H}).
Based on a consideration of timescales, \citet{2008arXiv0805.0801H} demonstrate that
these instabilities do indeed proceed much faster than global gravity, as required.

As simulations of the relevant processes
become more detailed, we are increasingly in a position to test the implications
of flow-driven molecular cloud and star formation. In this
paper we analyze numerical models published elsewhere (\citealp{2008ApJ...674..316H}, H\&08) 
to illustrate some general predictions of the flow-driven
picture, and help clarify some confusing issues about molecular
cloud and star-formation region lifetimes.

\section{Models}\label{s:techdetails}

\subsection{Overview}
The interstellar medium is filled with flows of various types, many of which result
in piling up material through shocks.  Recognition of this fact is at the heart of
the flow-driven cloud formation picture.  While many different types of initial conditions
can be envisaged -- flows driven by hot gas expansion (H II regions, supernova bubbles),
flows within mostly molecular gas -- we focus on the particular case of mildly supersonic
flows in atomic gas.  We do this for two reasons: first, it is computationally simplest for our
present situation; and second, it should be representative of conditions in the solar
neighborhood, where most of the gas is atomic (or ionized) and only a modest fraction
in mass and a very modest fraction in volume is present as molecular gas.  We expect
that many of the qualitative and even semi-quantitative results will apply to other
situations, though that needs further exploration.

We also chose to have equal uniform flows entering on either sides of our computational volume.
This is obviously not the most general case.  However, we have run some models in which
differing density atomic flows collide; this amounts to transforming into the frame
of the shock front of an expanding bubble (for example).  The results of these cases
do not differ substantially from the results we present here, and so 
we confine the discussion to the simplest possible cases.

We also chose to have smooth inflows and introduce only a modest perturbation of their
initial interface.  We did this not out of the absolute conviction that the flows themselves
are not turbulent, but with the desire of introducing as little substructure by hand
as possible.  By showing that even relatively smooth conditions rapidly introduce
turbulence, we can show that any other initial substructure will merely add additional
structure to our clouds.  

Finally, we chose not to use periodic boundary conditions.  This has absolutely crucial 
effects on the gravitational accelerations within the forming cloud. 
Global gravitationally-driven flows play important roles in cloud evolution, and these
{\em cannot} be seen in models with periodic boundary conditions.

By addressing the {\em formation} processes of a cloud, we are able to examine the initial
conditions which lead to structures and velocity fields important for star formation.

\subsection{Model Set}\label{ss:models}
We base our analysis and discussion on the models introduced in H\&08. They describe
the flow-driven formation of an isolated cloud with the parameters given in Table~\ref{t:modparam}.
All models are run on a fixed grid with the instreaming gas flowing along the 
$x$-direction, entering the domain (in opposing directions) at the two $(y,z)$-planes. 
The nominal resolution (i.e. the size of a single grid cell) is $8.6\times 10^{-2}$~pc. To trigger
the fragmentation of the (otherwise plane-parallel) region of interaction,
we perturb the collision interface. We chose the
perturbations of the collision interface from a random distribution of 
amplitudes in Fourier space with a top hat distribution restricted
between wave numbers $k=1..4$. 

The inflow is restricted to a cylinder of elliptical cross section with an
ellipticity of $3.3$ and a major axis of $80$\% of the (transverse) box size, 
mimicking two colliding gas streams in a more general geometry. 
Model Hf1 is a non-gravitating version of Gf1, to compare the role
of gravity versus that of the thermal instability for the fragmentation of the 
gas streams.

The inflow density in all models is $n_i=3$~cm$^{-3}$ at a temperature of
$T_i=1800$~K and an inflow velocity of $v_i=7.9$~km~s$^{-1}$, corresponding to
a Mach number of ${\cal M} = 1.5$. The flows are initially in thermal
equilibrium. The models start at time $t=0$ with the collision of the two
flows. The fluid is at rest everywhere except in the colliding cylinders.

\begin{deluxetable}{c|ccccc}
  \tablewidth{0pt}
  \tablecaption{Model Parameters\label{t:modparam}}
  \tablehead{\colhead{Name}&\colhead{$n_xn_yn_z$}
             &\colhead{$L_xL_yL_z$ [pc]}
             &\colhead{gravity}
             &\colhead{$t_{end}$ [Myr]}
             &\colhead{$\eta$ [pc]}}
  \startdata
  Hf1 & $256\times 512^2$ & $22\times 44^2$&no  & 14.5 & $2.2$ \\
  Gf1 & $256\times 512^2$ & $22\times 44^2$&yes & 14.5 & $2.2$ \\
  Gf2 & $256\times 512^2$ & $22\times 44^2$&yes & 14.5 & $4.4$
  \enddata
  \tablecomments{1st column: Model name. 2nd column: resolution. 3rd column:
                 physical grid size. 4th column: gravity. 5th column: end time of run.
                 6th column: amplitude of interface displacement.}
\end{deluxetable}

To solve the hydrodynamical equations, we used the 
higher-order gas-kinetic grid method Proteus
(\citealp{1993JCoPh.109...53P}; \citealp{1999A&AS..139..199S}; 
\citealp{2004ApJ...603..165H}; \citealp{2005MNRAS.356..737S};
\citealp{2006ApJ...648.1052H}, \citeyear{2007ApJ...665..445H}).
The code evolves the Navier-Stokes equations in their conservative 
form to second order in time and
space. The hydrodynamical quantities are updated in time-unsplit form.
Self-gravity is implemented as an external source term, also in time-unsplit form.
The Poisson-equation is solved via a non-periodic Fourier solver, using
the (MPI-parallelized) {\tt fftw} (Fastest Fourier Transform in the West) 
libraries. The heating and cooling rates are restricted to optically thin
atomic lines following \citet{1995ApJ...443..152W}.
Numerically, heating and cooling is implemented iteratively as a source
term for the internal energy $e$. While we do not include molecular line cooling,
the resulting effective equation of state reproduces the quasi-isothermal behavior
expected at high densities, reaching a temperature of $\approx 14$~K at 
densities of $\approx 10^3$~cm$^{-3}$.

The $x$-boundaries are partly defined as inflow-boundaries.
The inflow is defined within an elliptical surface in the $(y,z)$-plane.
The $y$ and $z$ boundaries, -- as well as the part of the $x$-boundaries
that is not occupied by the inflow --, are open, 
meaning material is free to leave the simulation domain through these boundaries.
For a detailed discussion of the numerical scheme and the models, see H\&08.

In view of the more general situation, we explored
a variety of geometries and flow parameters, such as expanding shells, unrestricted inflows,
and flows with different densities, the latter probably being the most general scenario for
e.g. swept-up clouds at the rims of supernova bubbles. Despite the variations, the basic
mechanisms and results do not change substantially. Specifically, the combination of
thermal and dynamical instabilities enhances local gravitational collapse in all cases.

\subsection{CO formation}\label{ss:comasses}
Our models do not explicitly follow the transition from atomic to molecular gas.
However, conditions in the dense post-shock gas should result in molecule
formation at some point
(see Glover \& Mac Low [\citeyear{2007ApJS..169..239G}, \citeyear{2007ApJ...659.1317G}] 
for a discussion of H$_2$ formation in turbulent flows).
To illustrate what would be observed as a molecular (CO) cloud
we make a simple approximation, motivated by the notion that CO formation 
requires shielding by dust grains.
For each time instance shown in \S\ref{s:movies}, 
we decide whether CO is ``present'' in a particular
grid cell by determining the expected radiation field integrated over solid angles.
If the effective UV extinction is equivalent to that of an angle-averaged
$A_V= 1$ and the local temperature is $T<50$~K, we assume CO is present in high abundance.
Because CO is rapidly dissociated at lower extinctions \citep{1988ApJ...334..771V},
we do not advect CO for simplicity.  
The radiation field at each grid point is calculated by measuring the incident radiation
for a given number of rays and averaging over the resulting sky. The ray number is determined
such that at a radius corresponding to half the box size, each resolution element is hit
by one ray. Thus, fine structures and strong density variations are resolved
(see also \citealp{2006MNRAS.373.1379H}). Since our CO maps are post-processed maps, there
is no timescale of CO formation involved, but CO is assumed to appear instantaneously in
sufficiently shielded regions. This simplification leads to ``more'' CO being present, countering
the neglection of advected CO. A full treatment of CO formation including
(even a simplified) chemical network (e.g. \citealp{2004ApJ...612..921B}), 
advection of chemical species and 
consistent radiative transfer would be a major computational challenge and beyond
the degree of detail necessary for our argument. The issue of CO formation is discussed further
in \S~\ref{ss:whatobs}. 

We concentrate on CO rather than on H$_2$ formation
because CO is the first available tracer to indicate the appearance of a molecular cloud, 
whereas H$_2$ is usually not directly observable in clouds. As Figures 1 and 3 of 
\citet{2004ApJ...612..921B} show, H$_2$ formation sets in well before CO formation 
due to self-shielding instead of dust shielding
-- specifically if traces of H$_2$ are already present in the swept-up material 
(e.g. \citealp{2001MNRAS.327..663P}). 

\section{At the Movies}\label{s:movies}

We now proceed to look at the evolution of the forming cloud with our crude treatment
of CO formation.  For reference, note that in the absence of any substructure or collapse,
the column density along the x-axis would vary with time as
\begin{eqnarray}
\langle N_H \rangle &=& 2 \times 7.9 \mbox{ km~s}^{-1} \times 3 \mbox{ cm}^{-3} t\\ 
                    &=& 1.5 \times 10^{20} \mbox{ cm}^{-2} (t/{\rm Myr})\,,\nonumber
\end{eqnarray}
and the mass of the cloud would be
\be
M_c \sim 500 \msun (t/{\rm Myr})\,,
\en
using the approximate area of our model cloud.
Thus, at the end of the three simulations ($t = 14.5$~Myr), the average column density
through the cloud in the absence of substructure would be $N_H \sim 2.2 \times 10^{21}$~cm$^{-2}$
or $A_V \sim 1.1$, and the cloud mass $M_c \sim 7500 \msun$.  The actual molecular mass
of the cloud will be smaller than this (see below), so our simulation is related to
the formation of a small molecular cloud.  

\subsection{Cloud Formation and Onset of Star Formation}\label{ss:cocloud}

\begin{figure*}
  \begin{center}
  \includegraphics[width=0.75\textwidth]{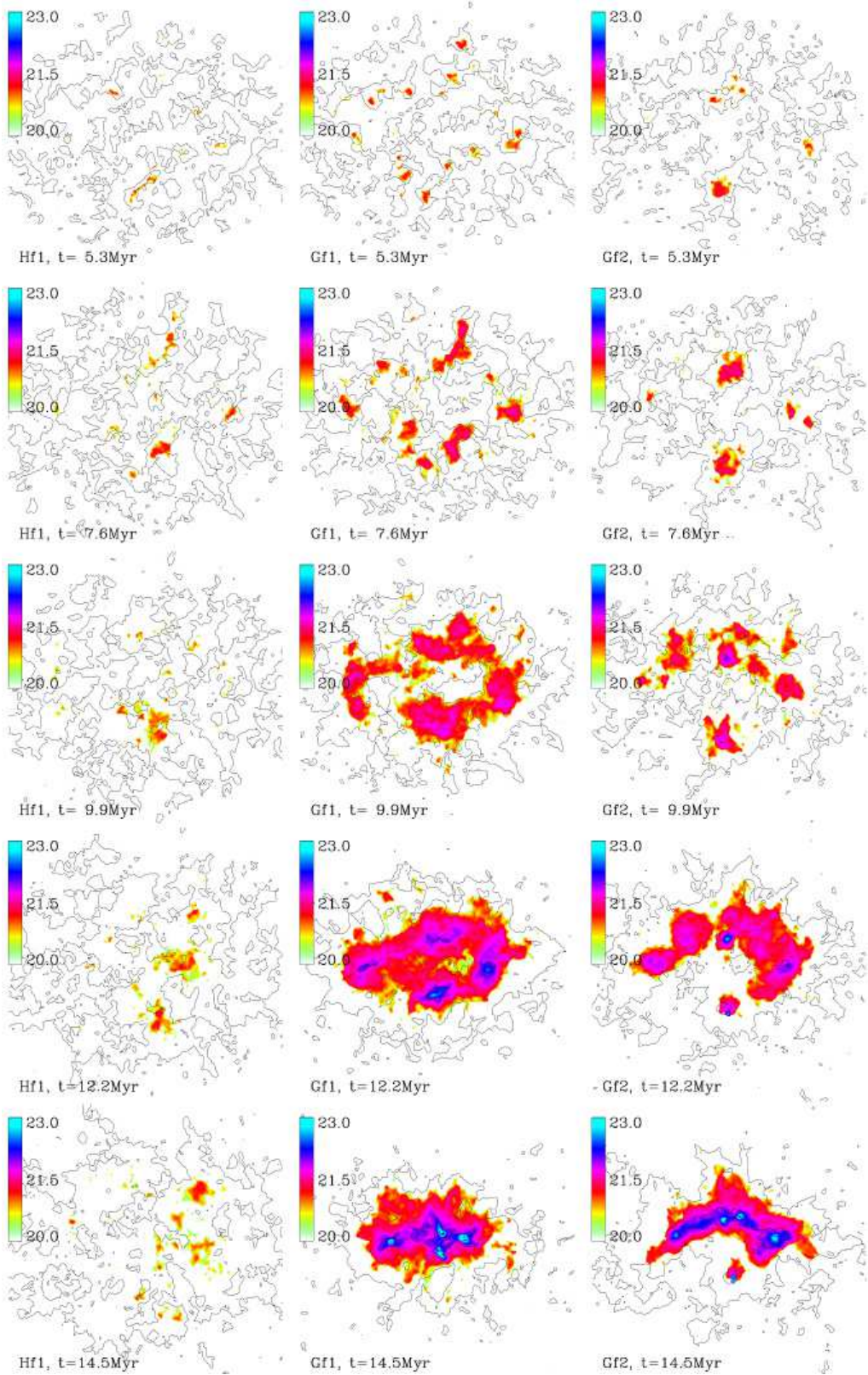}
  \end{center}
  \caption{\label{f:moviex}Time sequence of column density maps
           for models Hf1 (left), Gf1 (center) and Gf2 (right) 
           seen along the inflow. Each frame measures $22\times22$~pc$^2$.
           Contours show HI column density, at $\log N_{HI} = 20,21,22$, and
           colors show column density of gas containing CO in the range as indicated. 
           Self-gravity enhances the amount of CO.}
\end{figure*}

\begin{figure*}
  \begin{center}
  \includegraphics[width=0.75\textwidth]{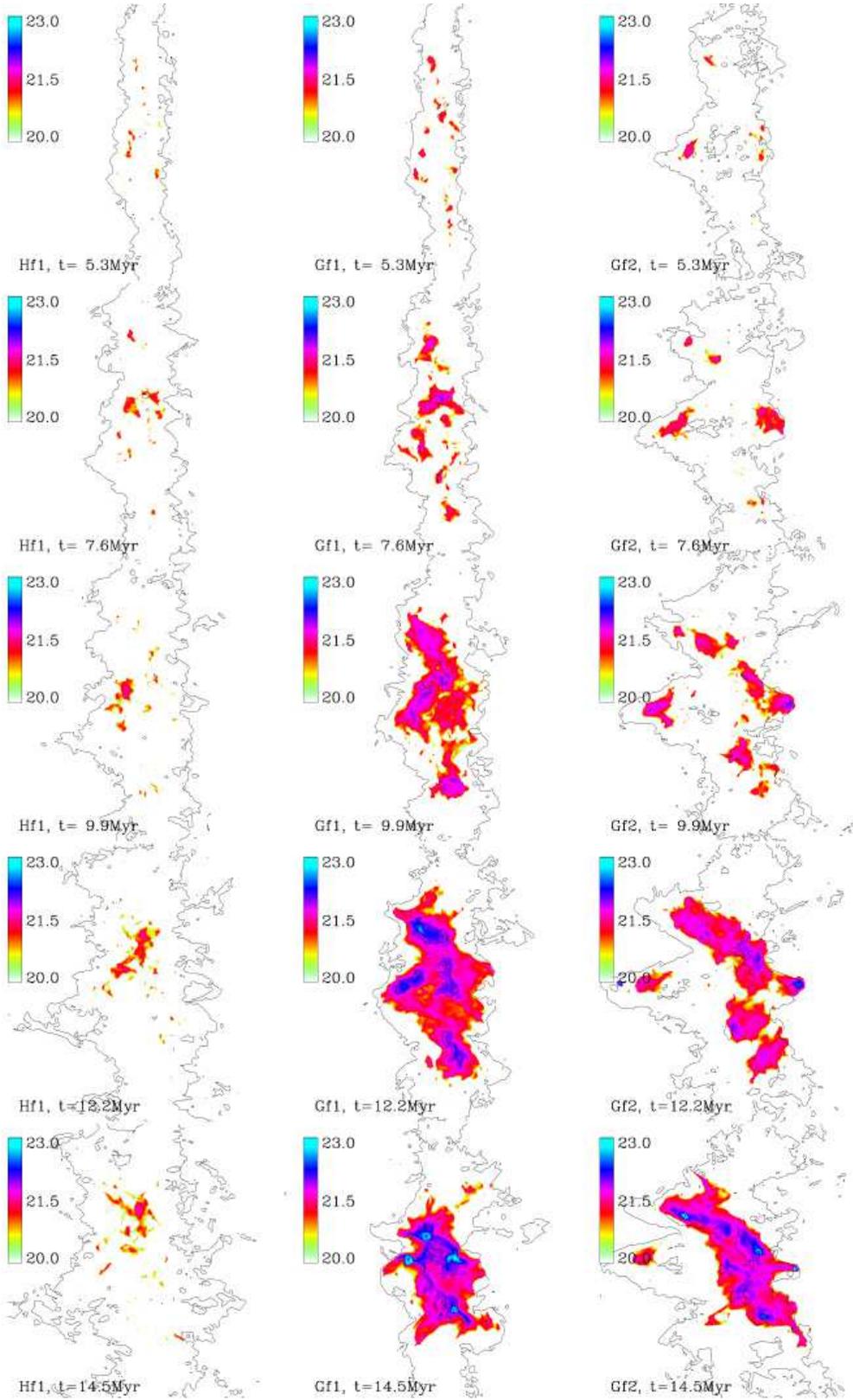}
  \end{center}
  \caption{\label{f:moviez}Time sequence of column density maps
           for models Hf1 (left), Gf1 (center) and Gf2 (right), 
           seen perpendicular to the inflow. Each frame measures $11\times 22$~pc$^2$.
           Contours show HI column density, at $\log N_{HI} = 20,21,22$, and
           colors show column density of gas containing CO in the range as indicated.}
\end{figure*}

Figures~\ref{f:moviex} and \ref{f:moviez} provide snapshots of the cloud evolution
at various times in the form of column density maps, where we use contours
to illustrate H I levels at levels of $\log N_{HI} = 20,21,22$, with colors
showing those hydrogen column densities at which CO is ``strongly present''.
Figure~\ref{f:moviex} shows the cloud formation seen along the inflows, and
Figure~\ref{f:moviez} gives an ``edge-on'' view. 

While at early times the small-scale structure in the HI column densities 
indicates the rapid fragmentation of the inflows due to thermal and dynamical 
instabilities in all three cases, the subsequent evolution is very different.  
Consider first the left-most columns in Figure~\ref{f:moviex},
which illustrates the evolution of model Hf1, in which gravity has been turned off,
seen along the direction of the flows.
Very little CO formation occurs over the $\sim 15$~Myr buildup of material.
The resulting ``cloudlets'' are small, relatively thin ($A_V \approx 1$), and in some
cases transient, due to the continuing input of turbulent energy from the inflows.

\begin{figure}
  \includegraphics[width=\columnwidth]{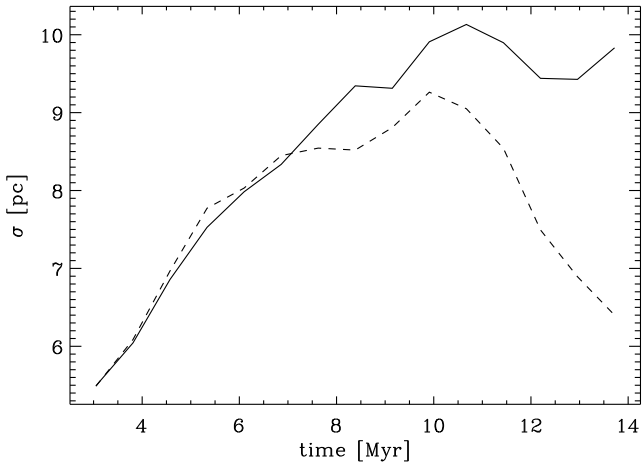}
  \caption{\label{f:profile}Mass profile width along the shorter (vertical in Fig.~\ref{f:moviex}) 
           axis of model clouds Hf1 (solid line) and Gf1 (dashed line) against time. Global collapse
           leads to a narrower width for model Gf1.}
\end{figure}

Things are dramatically different for the evolutionary sequence Gf1 in the middle column
of Figure~\ref{f:moviex},
where gravity is present at all times.  Although there is little change in behavior
at $t = 5.3$~Myr, by $t = 7.6$~Myr Gf1 has significantly more ``CO'', although
the concentrations are not strongly self-gravitating 
(\S\ref{ss:rapidcloud}).
By $t = 9.9$~Myr the model with gravity has formed a broken ring.
This structure results from the ``edge effect'' of non-linear gravitational
acceleration in a finite cloud (\citealp{2004ApJ...616..288B}; H\&08); it is in some sense
an ``echo'' of the initial cloud boundary.  
The slightly more compressed appearance of the HI contours (see also Fig.~2
of H\&08) indicate that it is the global gravitational modes and the resulting lateral
collapse motions (perpendicular to the inflow), which increase the local column density. 
The peak column densities at $10$~Myr are above $10^{22}$~cm$^{-2}$, compared to the 
average value $\sim 1.5 \times 10^{21}$~cm$^{-2}$ expected at that time.
Overall, the rapid production
of shielded molecular gas is strongly enhanced by {\em lateral} gravitational
collapse or sweep-up of gas, an effect initially suggested by 
Bergin \etal~(\citeyear{2004ApJ...612..921B}; see \S\ref{ss:rapidcloud}).
The lateral collapse overcomes the perceived limitations of the flow-driven cloud
formation scenario that within reasonable timescales the column densities would
be too low for efficient molecule formation (e.g. \citealp{2007ARA&A..45..565M}).
This can also be seen in Figure~\ref{f:profile}, which shows the density-weighted
width of clouds Hf1 and Gf1 along their short axis against time. At around
$8$~Myr, the widths start to separate, indicating the global gravity effects.
The effect in the widths is somewhat lessened by the edge effect due to the
non-linear gravitational accelerations, which nevertheless help sweeping up
material laterally and thus increasing the shielding. 
By $t = 12.2$~Myr (Fig.~\ref{f:moviex}) several local
small condensations appear which are very dense and strongly-self gravitating.
The overall region continues to collapse laterally under gravity until it forms
something close to a single filament at $t = 14.5$~Myr.  

The evolution of model Gf2 in the right-hand column is fairly similar to that
of Gf1.  The larger physical perturbations result in somewhat denser concentrations
of gas by $t = 9.9$~Myr, and the overall collapse tends to produce a single
narrow filament by the end of the simulation.

Figure~\ref{f:moviez} shows column densities in a side view of the three simulations.
The main concentrations formed in models Gf1 and Gf2 are driven initially at the
positions of inflections in the initial perturbed interface by dynamical focusing 
(\citealp{2003NewA....8..295H}; H\&08) due to the non-linear thin shell instability 
\citep{1994ApJ...428..186V}.
These views show that the resulting clouds and dense concentrations are far from
spherical or even spheroidal, but quite elongated and, eventually, strongly filamentary.

\begin{figure}
  \includegraphics[width=\columnwidth]{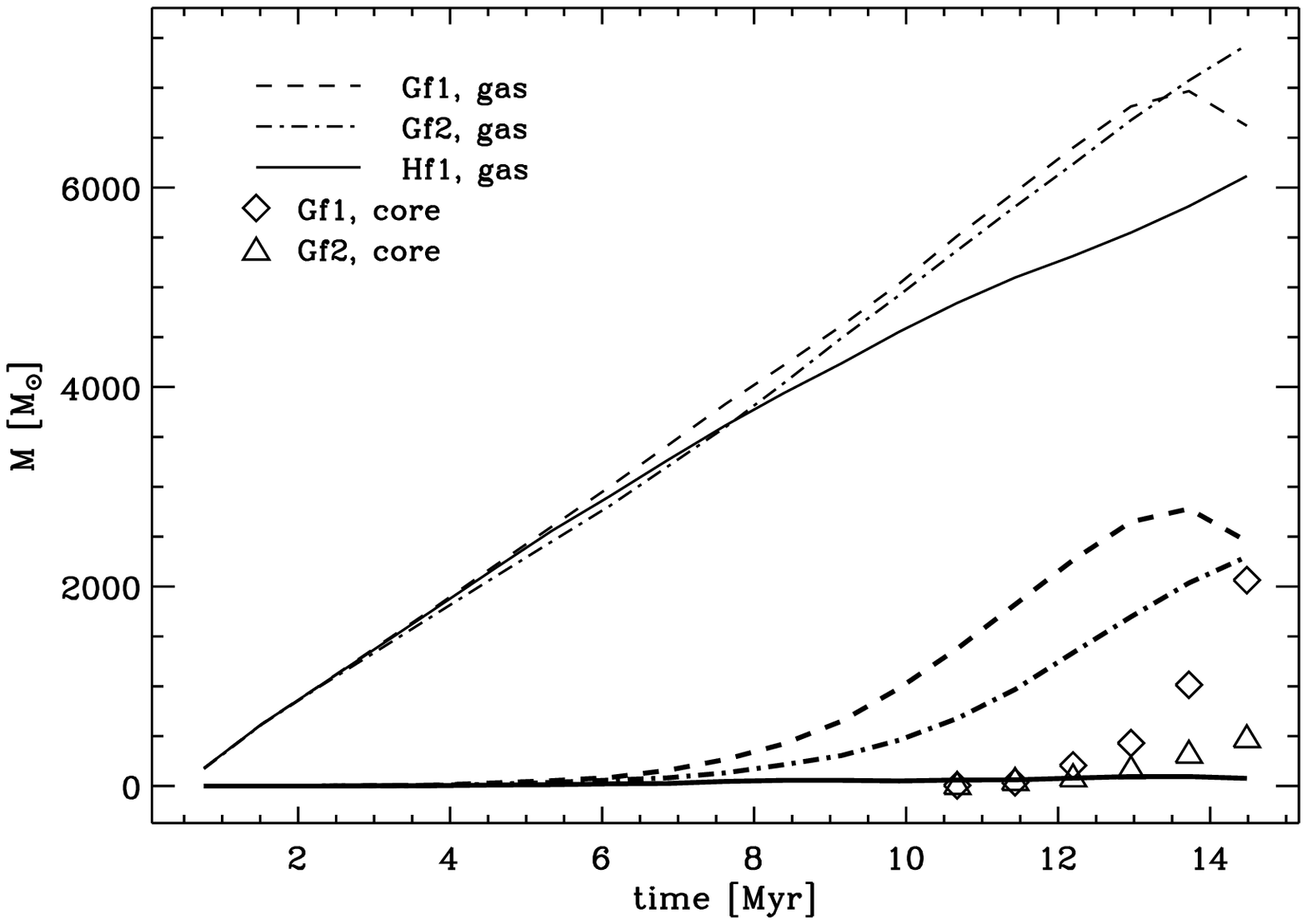}
  \caption{\label{f:comasses}Cloud masses in H~I (thin lines), CO (thick lines),
           and in all gravitationally bound cores (symbols) against time for models
           Hf1, Gf1 and Gf2. Approximately $2$ Myr after CO formation, the first
           cores form (see text for discussion). The turnover in the H~I and CO
           masses for model Gf1 is an artifact caused by switching off the cooling
           for densities $n>10^5$~cm$^{-3}$ (see Fig.~11 of H\&08 for a detailed explanation).}
\end{figure}

Once column densities of $N\approx 10^{22}$~cm$^{-2}$ have been reached (at $\approx 10$~Myr),
gravitationally dominated cores start to form (symbols in Fig.~\ref{f:comasses}).
Cores are identified by a modified version of CLUMPFIND \citep{1994ApJ...428..693W,2000ApJ...535..887K},
with the additional condition that  
the thermal plus kinetic energy content is less than half of the gravitational
potential energy (see H\&08).  
This is consistent, for example, with the results of \citet{1998ApJ...502..296O}, who studied
cores in Taurus with C$^{18}$O and found a column density threshold for star formation of about
$8 \times 10^{21}$~cm$^{-2}$.

\begin{figure*}
  \includegraphics[width=\textwidth]{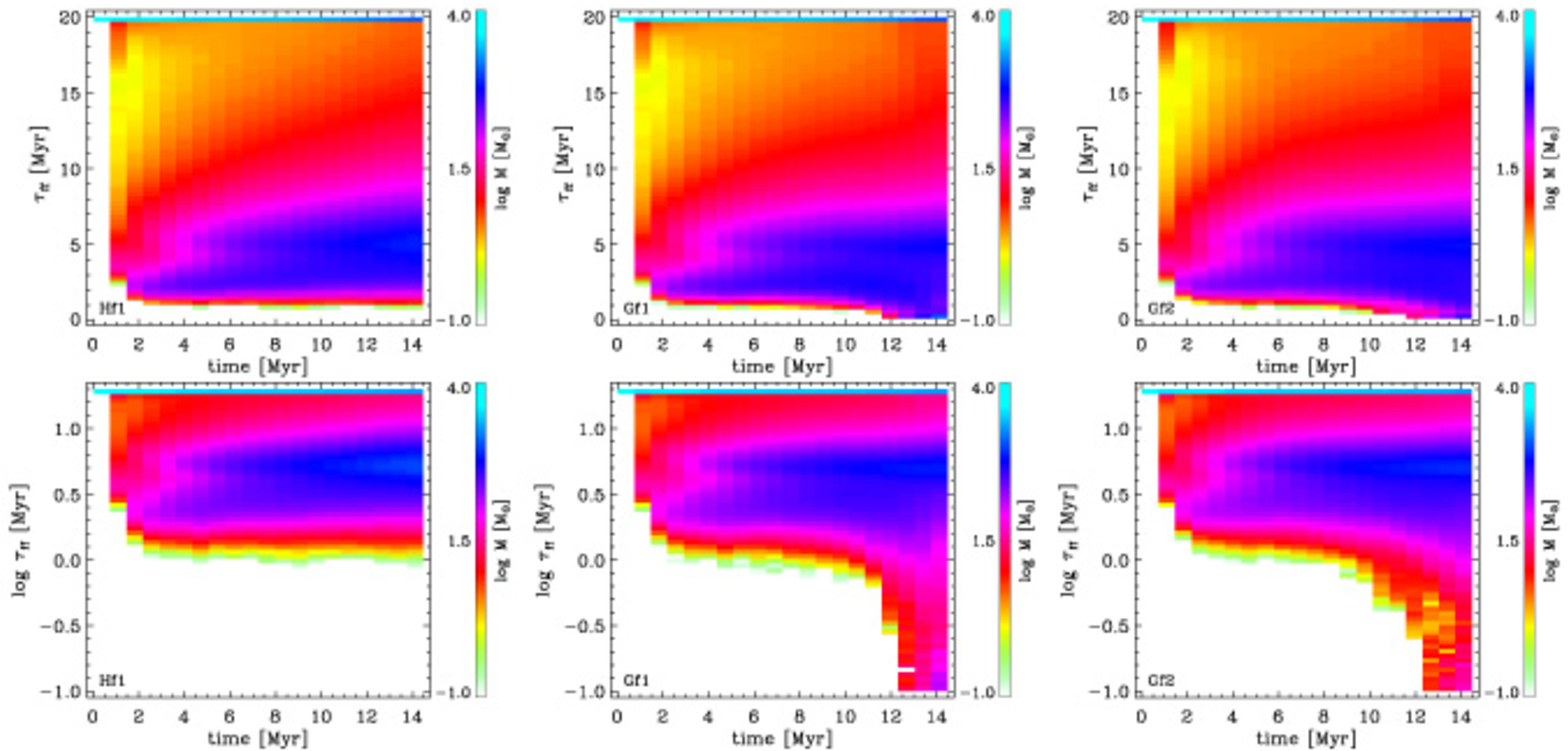}
  \caption{\label{f:taufftime}Time evolution of mass at given freefall time, for models
           Hf1, Gf1 and Gf2. {\em Top:} Linearly binned free-fall time, {\em bottom:}
           logarithmically binned freefall time. The bulk of the mass stays at freefall 
           times longer than the simulation time. At $t\approx 10$~Myr, a few $100$~M$_\odot$
           have reached freefall times $\tauff<1$~Myr, indicating local collapse. Even when
           logarithmically binned, the mass at small freefall times is sufficient to form
           stars.}
\end{figure*}

We cannot follow the collapse of these dense regions to anything
approaching the sizes of stars.  However, it is instructive to consider not only whether
the regions are self-gravitating but the timescales upon which they might collapse.
Figure~\ref{f:taufftime} shows the distribution of mass as a function of its local
free-fall timescale, binned linearly for the top row and 
logarithmically for the bottom row for clarity.  The bulk of the mass 
in the simulation domain remains at long free-fall times.  The formation of the first
gravitationally-bound cores occurs in the simulations with
gravity at about 10 Myr, as shown in Figure \ref{f:comasses};  
at about the same time, a small amount of mass exhibits free-fall timescales of order
1~Myr or so.  It is only at $t \sim 11-12$~Myr that appreciable amounts of mass 
are present with free-fall timescales of $< 1$~Myr; thus, we estimate that this is
the epoch at which star formation begins.  It is evident, given the increase in
the number of self-gravitating cores and the decrease in free-fall times, that
star formation in this model would be an increasing function of time to the end
of the simulation.

\section{Discussion}

\subsection{What would we observe?}\label{ss:whatobs}

The cloud evolution that we have presented obviously takes far too long to be 
observable.  For purposes of comparison with observation one can think 
in terms of observing four different clouds, each having the same evolution, but studied
at different epochs.  We also compare at each stage with some observed clouds
of similar properties, with the primary goal of checking whether our prescription
for CO formation appears reasonable, either with respect to observations or to
theoretical treatments for these clouds.  We restrict attention to model Gf1 initially.

``Cloud 1'' (the top row of Figure \ref{f:moviex}) would be difficult to detect
in molecular gas emission, but could be inferred through absorption line studies and
extinction, especially in the small blobs.  This object would be
classified as a diffuse H I cloud with a visual extinction $A_V \sim 0.4$. 

``Cloud 2'' (second row), with an age of 7.6~Myr, would 
still be classified as a diffuse H I cloud.  It has a small amount of CO, mostly in 
localized positions; Figure \ref{f:comasses} demonstrates that very little of
the cloud is CO-bearing at this phase.  
The average extinction through the cloud would be $A_V \sim 0.6$,
with of course some localized higher column density regions; such properties are
roughly consistent with local lines of sight to, for example, $\zeta$~Oph and
$\zeta$~Per, for which van Dishoeck \& Black's (\citeyear{1988ApJ...334..771V}) 
models are consistent with the presence of small amounts of CO.

Here we should emphasize that our prescription for CO formation relies heavily on results
such as those of Figure 5a of \citet{1988ApJ...334..771V}, which indicate
photodissociation timescales $>1$~Myr for $A_V \gtrsim 1$.  Their analysis implies the
presence of substantial amounts of H$_2$ - roughly comparable to the amount of H in the
$\zeta$~Per and $\zeta$~Oph lines of sight.  However, our models would not form any
significant amounts of H$_2$ at the stage of ``Cloud 2'', 
because essentially none of the gas is at densities
high enough to form molecular hydrogen rapidly enough
(\citealp{2004ApJ...612..921B}; formation timescales $>1$ Myr at 
densities $< 10^3$~cm$^{-3}$, corresponding to local free-fall timescales $>1$~Myr; 
see Figure \ref{f:taufftime}).  The only way that there would be significant amounts
of H$_2$ present at this time (or in ``Cloud 2'') is if it were present in the inflowing
gas (e.g., \citealp{2001MNRAS.327..663P}).  Thus, if anything we may have overestimated the
amount of CO present at this stage.

``Cloud 3'' (third row) now begins to show significant amounts of CO.  According to 
Figure \ref{f:comasses}, the mass in CO is now about $500 M_\odot$ in the small-perturbation
model Gf1 and about $10^3 \msun$ in the large-perturbation model Gf2.  
However, no self-gravitating cores have formed at this point.

One might regard ``Cloud 3'' as an example of the most massive of the so-called translucent
clouds, which are generally observed at high galactic latitude 
(e.g., \citealp{1985ApJ...295..402M}; \citealp{1996ApJS..106..447M}; \citealp{2003ApJ...592..217Y}).
The constraint on galactic latitude probably
represents an observational selection effect rather than a complete absence of
such objects in the galactic plane.  Observationally, star formation is rare or absent
in this type of cloud, consistent with our simulations.
For example, in their study of the H{\small I} filament associated with the molecular
complex including MBM 53, 54, and 55, which has a mass of about $1200 \msun$ and
a length of about 30~pc (assuming a distance of 150~pc), \citet{2003ApJ...592..217Y}
found no evidence for any star-forming cores, though there is evidence for one
or two young weak-emission T Tauri stars in the region.
Note also that translucent clouds are known to have variations in CO abundances
of an order of magnitude, even for various positions within the same cloud
(\citealp{1998ApJ...504..290M}, and references therein) which of course is what would
be expected for a turbulent, structured translucent cloud like ``Cloud 3''.
 
``Cloud 4'', which appears at $t \sim 12$~Myr, is now a molecular
cloud, with a mass associated with CO of
order 1200 $\msun$.  It now contains a small but significant amount of mass which is
both gravitationally bound and is at densities corresponding to free-fall timescales
less than 1 Myr; thus we consider that it has now begun to form (low-mass) stars.

Finally, ``Cloud 5'' is now a full-fledged molecular cloud, with a mass associated with
CO of order 2500 $\msun$ and dense cores comprising of
order at most 25\% of the mass associated with CO and of order 5\% of the total
mass of the cloud including H I.  The cores are contained within a strongly confined
region spatially, due to the global gravitational collapse of the material.

The behavior of model Gf2, with a larger initial perturbation, is very similar,
except that high densities are achieved a little earlier, the cloud collapses into
a better-defined filament, and the efficiency of massive core/star formation is
larger (Figure \ref{f:comasses}).

Note that the fraction of mass in gravitationally-bound regions does {\em not}
directly correspond to the rate of star formation.  For example, in model Gf2
at the end of the simulation there is about $2000 \msun$ in gas containing
CO, but only $\sim 10^2 \msun$ of this material is actually contained in regions
with free-fall times of 1 Myr or less (Figure \ref{f:taufftime}).  Thus in all
cases we expect star formation efficiencies would be on the order of a few
percent at most, by the end of the simulations.

The sizes and masses of models Gf1 and Gf2 at the final, ``Cloud 4'' stage,
are reasonably consistent with those of one of the major filaments in the
Taurus molecular cloud \citep{2002ApJ...578..914H}.  One may even take this a step further
and argue that star (core) formation could have only started in these models
at times $\sim 11-12$~Myr, so that at 14.5 Myr the age spread is only of order
2-3 Myr, again reasonably consistent with that of Taurus \citep{2003ApJ...585..398H}; but
one would then need to shut off star formation to maintain the agreement with
observation (\S 4.6).

The above sequence of ``clouds'' is not meant to suggest that all low-mass 
atomic clouds will eventually become molecular, star-forming regions.
In particular, high-latitude clouds may well remain atomic, simply because
they cannot sweep up enough mass.

\begin{figure}
  \begin{center}
  \includegraphics[width=0.9\columnwidth]{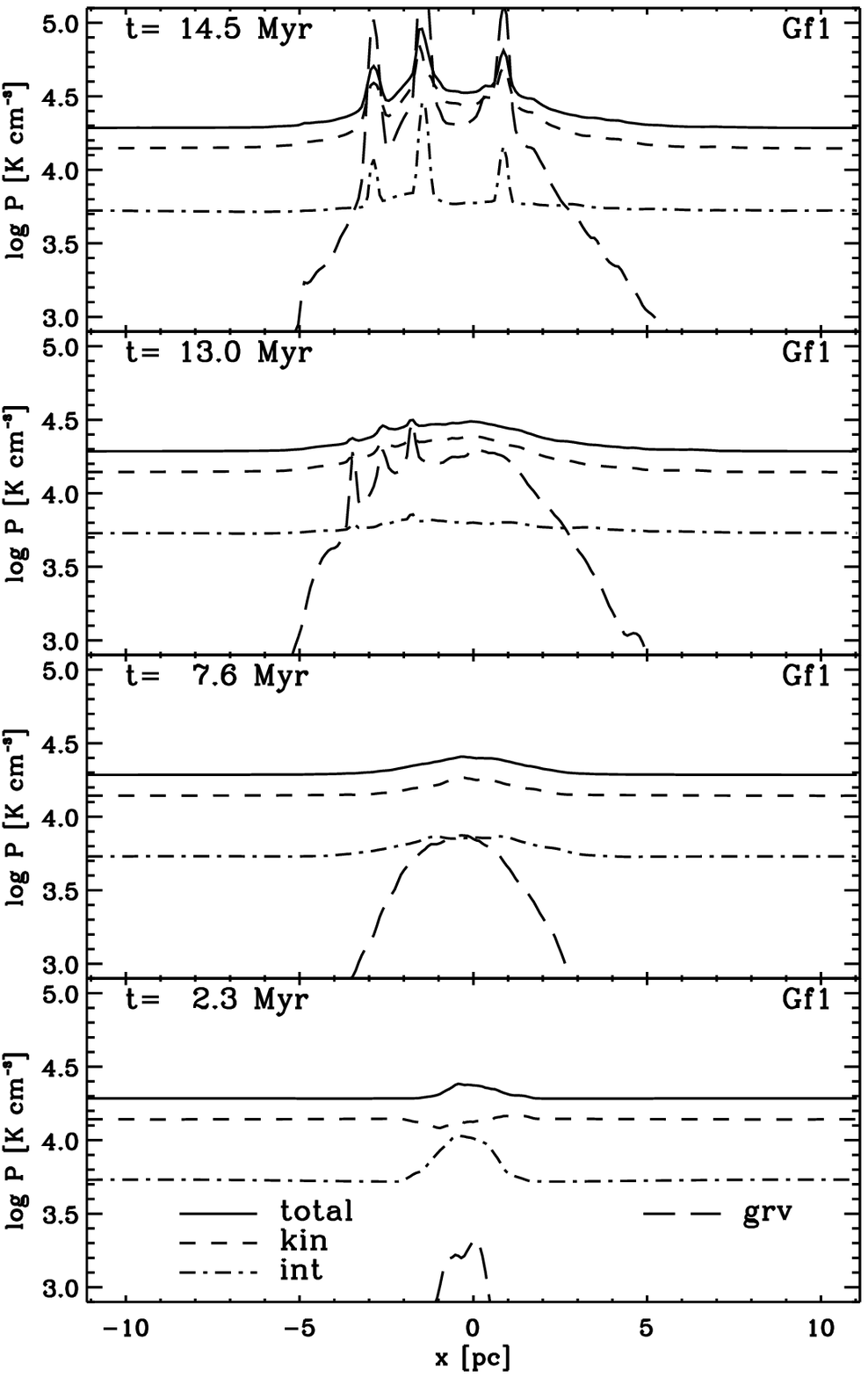}
  \end{center}
  \caption{\label{f:pressprof}Pressure profiles along the inflow axis for model Gf1.
           Shown are total, dynamical (``kin''), thermal (``int'') and gravitational (``grv'')
           pressure.}
\end{figure}

\subsection{Rapid molecular cloud formation}\label{ss:rapidcloud}

The simulation sequences shown in Figures \ref{f:moviex} and \ref{f:moviez}
support previous suggestions of rapid cloud and star formation.
Specifically, HBB01 argued that the paucity of substantial molecular clouds without
star formation meant that collapse of molecular cloud cores must follow soon
after cloud formation.  They further suggested that this was the result of
the required shielding column density for molecular gas against the dissociating
interstellar radiation field, which was similar to that needed for gravitational
collapse.  Our simulations demonstrate this behavior, with molecular (CO) gas setting in
near $t \sim 10$~My and core formation soon after, with filamentary gas structure.

A few principles explain
why our simulations show the observationally required behavior. 
One is that the cloud is formed with relatively
high densities even in the atomic phase, and thus is ``pre-formed'' with a short free-fall
time making rapid collapse possible.  Figure~\ref{f:pressprof} shows that
the pressure profiles along the inflow axis, averaged over
the surface area of the cloud for model Gf1,
roughly reflect the ram pressures of the inflows
until gravity becomes important.  This means that the density within
the cloud will approach
\be
\rho \sim \rho_i v_i^2 (k_B T /\mu m_H)^{-1}\,,
\en
where $\rho_i$ and $v_i$ are the inflow density and velocity, respectively,
and $T$ is the temperature.  In terms of densities and velocities
$\tilde{\rho}_i$ and $\tilde{v}_i$ relative to the values used in these
simulations ($v_i=7.9$~km~s$^{-1}$ and $n_i=3$~cm$^{-1}$, see \S\ref{ss:models}),
\be
\rho \sim 1.3 \times 10^{-21} T_{30}^{-1} \, \tilde{\rho}_i \tilde{v}_i^2 {\rm g}\, \mbox{cm}^{-3} \,,
\en
or
\be
n \sim 750 T_{30}^{-1} \, \tilde{\rho}_i \tilde{v}_i^2 \, \mbox{cm}^{-3} \,,
\en
where $T_{30}$ is the temperature measured in units of 30 K, typical of the cold neutral
material.  (Note that in the simulations the mean molecular weight is assumed to be unity,
as used in these equations.) The free-fall time is then
\be
\tauff = \left ( {3 \pi \over 32 G \rho } \right )^{1/2}
\sim 1.9 T_{30}^{1/2} \, \tilde{\rho}_i^{-1/2} \tilde{v}_i^{-1} {\rm Myr}\,.
\en
This estimate is in reasonable 
agreement with Figure \ref{f:taufftime}, which shows a substantial amount of
gas at $\tauff \sim 2$~Myr, with higher temperature material at $\tauff \sim 6$~Myr. 
Thus, simply due to ram pressure, moderately cold gas in the cloud is formed at
densities such that free-fall timescales are short.  This allows rapid gravitational
collapse once sufficient mass has been accumulated.  

Another feature of our model is that, as
\citet{2004ApJ...612..921B} showed using a one-dimensional shock model with
chemistry, much of the time spent accumulating cloud mass
is spent in the atomic phase; such clouds would not be recognized as ``proto'' molecular
clouds.  Beyond this, however, local gravitational collapse
tends to occur with CO formation because global gravitational collapse is responsible for 
accelerating the formation of high-column density, highly-shielded gas which can become molecular.
This was already foreseen by Bergin \etal, who  
suggested that {\em lateral} collapse under gravity would be responsible for producing
``runaway'' increases in column density, enhancing the rate of molecule formation
beyond what can be achieved in a one-dimensional flow.
We observe precisely that behavior in our cloud sequence in Figures \ref{f:moviex}
and \ref{f:moviez} (compare the left-hand column without gravity to the center and
right-hand columns with gravity). 
\citet{2008arXiv0806.4312D} discuss the formation of H$_2$ clouds in spiral arms and
argue that substantial fractions of H$_2$ can be formed rapidly even without self-gravity.
This does not contradict our findings, since -- as discussed in \S\ref{ss:comasses} --
H$_2$ is expected to form well before CO, at lower column densities. 
In addition, the densities of the clouds found in these large-scale simulations are
generally smaller than typical molecular clouds in the solar neighborhood.  Finally, 
even if H$_2$ can form without self-gravity, this does not mean that both H$_2$ {\em
and} CO will not form faster with self-gravity.

\citet{2004ApJ...616..288B} showed that cold sheets were also susceptible to global
gravitational collapse.  Ignoring concentrations due to edge effects, which can
occur much faster, they found that the timescale for
global sheet collapse with outer radius $r$ is 
\be
t \sim \left ( {r \over G \Sigma} \right )^{1/2}\,.
\en
If we ignore the collapse motions and simply allow mass addition at a rate
such that $\Sigma = \rho_i v_i t$, then we can estimate the distance scales over
which global collapse should be present as
\be
r \sim 5.4 t_7^3 \, \tilde{\rho}_i \tilde{v}_i \, {\rm pc}\,,
\en
where $t_7$ is the time measured in units of 10 Myr.
Thus it is not surprising that global gravitational collapse is strongly underway
at the time of massive core condensation in the simulations.  

\citet{2004ApJ...616..288B} also demonstrated
that finite sheets experience highly non-linear gravitational 
acceleration as a function of position which tends to cause material to
pile up near cloud edges.  Our simulations confirm this;
even though the resulting clouds are three dimensional instead of being
sheets, they are flat enough that the initial elliptical boundary is echoed
in the resulting filamentary collapse (an even stronger version of the edge
effect is present in \citet{2007ApJ...657..870V}, whereas we took steps to suppress the strength
of the edge effect by reducing the inflow mass addition near the boundaries).

To summarize, reasonable ram pressures of inflowing material result in cold atomic
gas with densities sufficiently large that gravitational collapse can be rapid once
sufficient material is accumulated.  Substantial turbulent fragmentation is needed to
form stellar mass objects.  Global collapse is generally important in driving up column
densities and enhancing the formation of filamentary structures through edge effects.
Free-fall times, Jeans masses and lengths will decrease, and global collapse will occur 
faster for higher inflow rates and ram pressures, which probably are needed to make more
massive clouds than those of our simulations.

We emphasize that simulations with periodic boundary conditions cannot capture the
evolution seen here, in particular the rapid increase in column density within
filamentary structure.  Simulations with periodic gravity can only be relevant if started
with substantial perturbations, and limited to scales much smaller than the overall
sizes of clouds.

\subsection{``Accelerated'' Star Formation}\label{ss:accelsf}

Palla \& Stahler (\citeyear{2000ApJ...540..255P}, \citeyear{2002ApJ...581.1194P}) 
estimated ages for stars in several local star-forming regions,
finding that the rate of star formation generally showed an increase up to the last
1 Myr or so.  As Hartmann~(\citeyear{2002ApJ...578..914H}, \citeyear{2003ApJ...585..398H})  
pointed out, it is implausible that all of
these star-forming regions should be coordinated in their evolution; it 
must be that the typical lifetimes of star-forming regions are at most a few Myr.
Hartmann~(\citeyear{2002ApJ...578..914H}, \citeyear{2003ApJ...585..398H})
also pointed out some problems 
related to neglect of observational errors, logarithmic vs. linear binning in age,
and likely isochrone or birthline problems which exaggerate the age spreads of these
regions.  Nevertheless, there is evidence that at a small fraction of the stars
associated with these areas are a few Myr older than the bulk of the population.

Accelerated star formation is expected in a trivial sense, since starting from a rate
of zero star formation to a finite star formation rate mathematically requires 
an acceleration.  However, it also occurs naturally in our evolutionary model of
molecular cloud formation.  The Gf1 and Gf2 simulations shown in Figures \ref{f:moviex}
and \ref{f:moviez} show that a few dense concentrations show up by 10 Myr or so
(see also the core mass history, Fig.~5 in H\&08).
One can easily infer that, if we had sufficient resolution, we would observe a few
stars being formed from the very densest of the dense concentrations first; later on,
as global and local collapse proceeds, more and more stars would form.  

Our simulations emphasize that the global age spread in a star-forming region is
an upper bound to the timescales of local collapse, as compressions will never be
perfectly coordinated in time across any given cloud.

\subsection{The Onset of Star Formation}\label{ss:onsetsf}
In our model, turbulence generated through the cloud formation
process provides a mechanism to produce 
thermal and dynamical fragmentation. While this fragmentation is highly efficient in 
generating cold high-density cloudlets, the ``pre-formation'' of such cloudlets
is limited by the minimum Jeans mass achievable.
With $c_s$ denoting the (isothermal) sound speed, we have  
\be
M_J = \left ( {\pi c_s^2 \over G } \right )^{3/2} \rho^{-1/2}
\sim 570 T_{30}^2 \tilde{P}_{ram}^{-1/2} \, \msun\,,
\label{e:mjeans}
\en
where $\tilde{P}_{ram} = \tilde{\rho}_i \tilde{v}_i^2$, again in units of our simulations.  
To form e.g. solar-mass objects, higher pressures are needed. Since invoking
``supersonic turbulence'' as a fragmentation mechanism in its own right without specifying
its physical source is somewhat unsatisfactory, the only remaining free energy source leading
to a self-consistent picture of the formation of pre-stellar cores is gravity
(e.g. \citealp{1992ApJ...395..140B}; \citealp{2008MNRAS.385..181F}).

\begin{figure}
  \begin{center}
  \includegraphics[width=0.9\columnwidth]{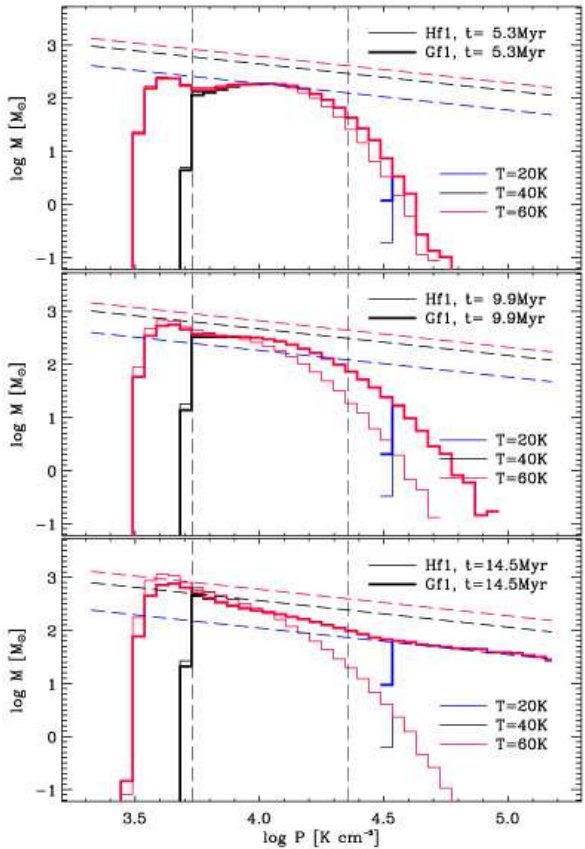}
  \end{center}
  \caption{\label{f:moviep}Mass (in $M_\odot$) within logarithmic pressure bins
           for models Hf1 and Gf1, at the same times as in e.g. Figure~\ref{f:moviex}.
           Colors indicate temperature thresholds. The left vertical black dashed line
           stands for the thermal pressure of the inflows, while its right-hand counterpart
           shows the ram pressure of the inflows. The slanted lines in the temperature color
           code indicate the Jeans mass (eq.~\ref{e:mjeans}). With increasing time, gravity
           drives more and more mass to higher pressures, allowing the fragmentation into
           successively smaller structures \citep{2008MNRAS.385..181F,2008ASPC..387..240V}.}
\end{figure}

We have already seen that gravity helps along the ``CO formation'' in our clouds (Figs.~\ref{f:moviex} and
\ref{f:moviez}), and also that unless the simulations include gravity, they are not going to form 
the high-density cores with freefall times $\tauff<1$~Myr (Fig.~\ref{f:taufftime}). 
Figure~\ref{f:moviep} demonstrates that it is gravity which eventually leads to the high pressures
required for fragmentation into small masses. It shows the mass-weighted pressure histograms of
models Hf1 and Gf1 in the same time sequence as in e.g. Figure~\ref{f:moviex}. At $t=5.3$~Myr, 
the distributions are indistinguishable. With increasing time, gravity drives more and more mass
to higher pressures, until the pressure distribution follows the Jeans mass (eq.~[\ref{e:mjeans}])
at the lowest temperature threshold (indicated by the blue dashed line). 
In contrast to model Gf1, the pressure distribution of model Hf1 stays more or less constant with time.

\subsection{Turbulent Support}\label{sss:turbsupp}
\begin{figure}
  \includegraphics[width=\columnwidth]{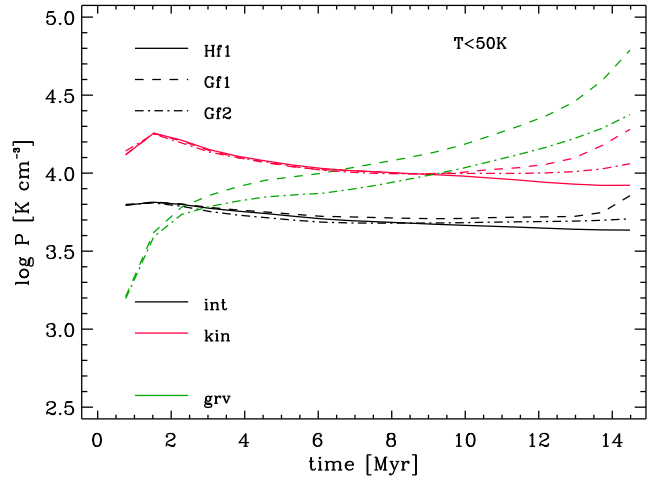}
  \caption{\label{f:energies}Energy densities (or pressures) of the three isolated
           cloud models Hf1, Gf1 and Gf2, for gas at $T<50$~K, approximately corresponding to
           the molecular gas. The internal energy
           (black lines) stays constant, while the gravitational energy (green lines) 
           increases with time, offsetting the slight decrease in kinetic energy (red lines).}
\end{figure}

\begin{figure*}
  \includegraphics[width=\textwidth]{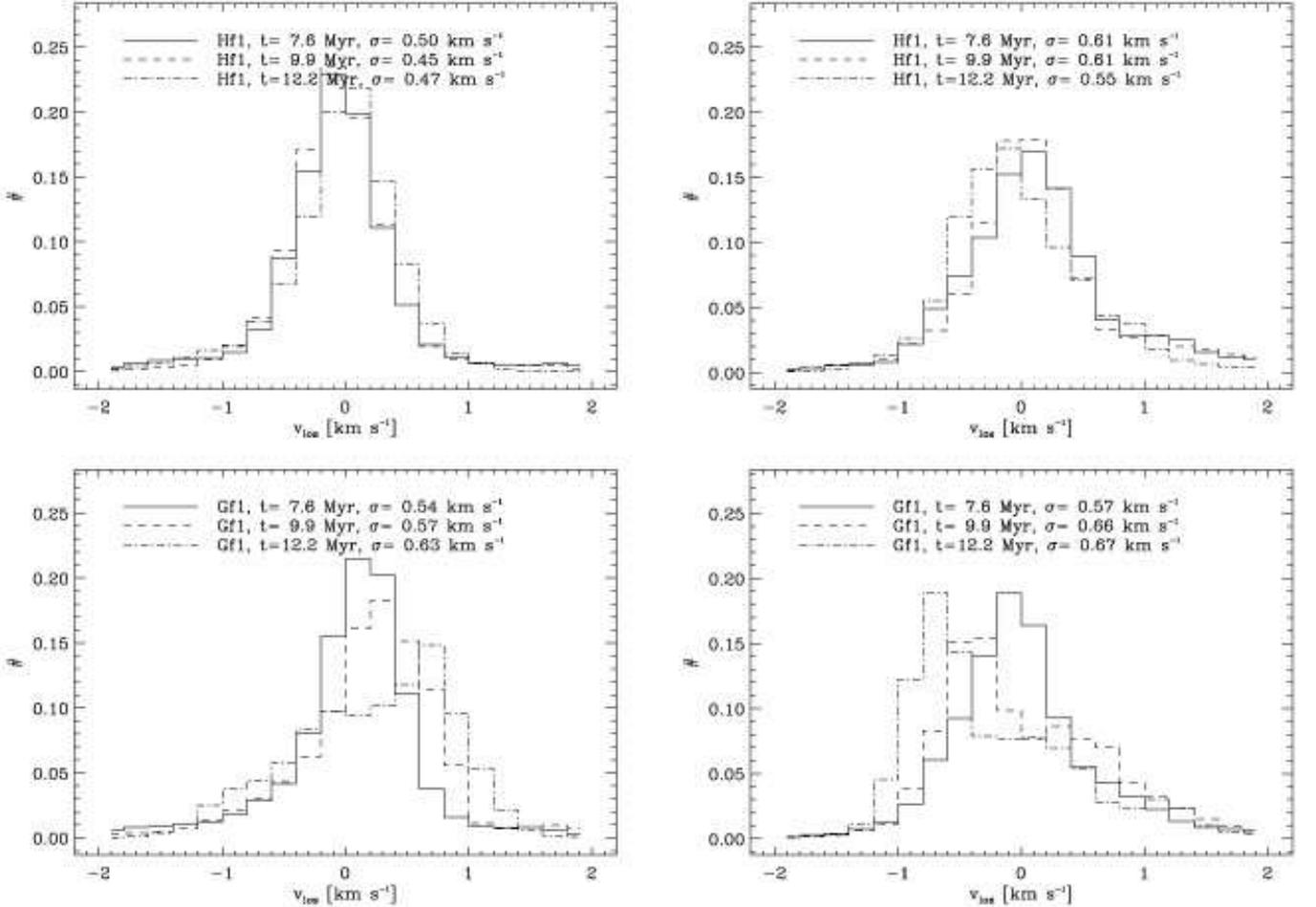}
  \caption{\label{f:losvd}Line-of-sight velocity distribution for models 
           HF1 (top) and Gf1 (bottom), for a single line of sight parallel to the 
           shorter ({\em left}) and longer ({\em right}) 
           axis of the cloud perpendicular to the inflow. 
           Three times as in Figure~\ref{f:moviex}.}
\end{figure*}

It has been claimed that the loss of turbulent support (e.g. \citealp{2007ApJ...666..281H})
would lead to an (accelerated) star 
formation in molecular clouds \citep{2000ApJ...540..255P,2002ApJ...581.1194P}. 
A closer look at the energy budget 
(Fig.~\ref{f:energies}) in our model clouds however reveals that in the context of flow-driven 
cloud formation, it is actually the build-up of mass and thus the deepening of the 
gravitational potential well which leads to collapse, but not the loss of turbulence (see also 
Fig.~8 of \citealp{2007ApJ...657..870V}). The energy densities were calculated using only gas
at $T<50$~K, approximately corresponding to the molecular (Gf1, Gf2), but at least cold, 
dense gas (Hf1). The 
gravitational energies grow faster with time than the kinetic energies drop -- and eventually,
once the molecular cloud has ``appeared'' (see Figs.~\ref{f:moviex} and \ref{f:moviez}), 
the kinetic energies start to rise, indicating that the turbulent ``motions'' at that stage 
are driven to some extent by gravitational collapse (see also \citealp{2006MNRAS.372..443B};
 \citealp{2007ApJ...657..870V}; \citealp{2008MNRAS.385..181F}; 
\citealp{2008ASPC..387..240V} for a recent discussion). In other words, while the 
turbulence might decay slightly during the formation of the model cloud (in the atomic phase), 
it is {\em increasing} -- driven by gravity -- during the molecular phase. 
One would expect stellar feedback to even strengthen this trend, if anything.
Figure~\ref{f:losvd} demonstrates that despite the gravitational dominance, the linewidths
still can look ``turbulent''. The linewidths shown were calculated for all gas at $T<50$~K, 
along a single line-of-sight of 16 resolution elements width through the center of the cloud.

We note that the kinetic energies shown in Figure~\ref{f:energies} are subvirial by a factor
of up to $3$ at late times, and that the linewidths (Fig.~\ref{f:losvd}) 
are on the smallish side. We believe this
is due to the combination of three issues. First, the head-on colliding flows do not allow
for large-scale shearing motions. Those would be expected e.g. in molecular clouds forming
in spiral arms. Numerical models of this process \citep{2006MNRAS.365...37B} including a shear
component reproduce turbulent linewidth-size relations \citep{1981MNRAS.194..809L}.
Second, non-uniform inflows will increase the level of turbulence in the resulting cloud, 
as demonstrated by \citet{2007MNRAS.374.1115D}. They found the slope of the 
linewidth-size relation to flatten with increasing filling factors of the clumpy inflows.
Third, the dynamical range for gravitational collapse in our models is somewhat limited, 
possibly suppressing turbulent motions once the cloud has collapsed.

\begin{figure}
  \includegraphics[width=\columnwidth]{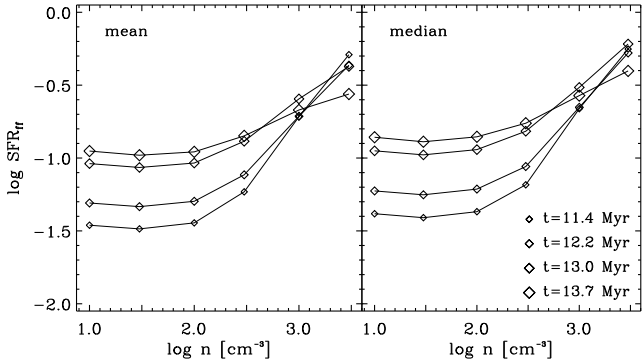}
  \caption{\label{f:krumholztan}The star formation rate normalized by the free fall time
          $SFR_{ff}$ (eq.~[\ref{e:bogussfr}]). The plot has been generated following
          the recipe given by \citet{2007ApJ...654..304K}, i.e. the free-fall time is computed for the 
          mean ({\em left}) and median ({\em right}) density above the density threshold given
          on the $x$-axis. Symbol sizes indicate simulation times.}
\end{figure}

We take the discussion of the perceived loss of turbulent support as an opportunity to 
demonstrate how the star formation in our models 
(strictly speaking, we are only forming cores)
would be represented by the 
normalized ``star formation rate'' of \citet{2005ApJ...630..250K} and \citet{2007ApJ...654..304K},
\begin{equation}
SFR_{ff}\equiv \frac{\dot{M}}{M(>n)}\,\tauff.\label{e:bogussfr}
\end{equation} 
Here, $M(>n)$ is the mass of the cloud above a density threshold $n$, $\dot{M}$ is the 
star formation rate, and $\tauff$ is the freefall time at a given density. The choice of 
this density obviously will affect the results. Figure~\ref{f:krumholztan} shows the
$SFR_{ff}$ as defined above, for various times against the threshold density $n$ (model Gf1).
As a proxy for the star formation rate $\dot{M}$ we use the time derivative of the 
mass accretion history (see e.g. symbols in Fig.~\ref{f:comasses}). This includes the formation
of new cores as well as the mass accretion onto already formed cores. It is also consistent with
the estimates of $\dot{M}$ from the simulations discussed by \citet{2007ApJ...654..304K}.

At early times ($t=11.4, 12.2$~Myr), the models are consistent with the analytical predictions
(see Fig.~5 of \citealp{2007ApJ...654..304K}): the $SFR_{ff}$
ranges around a few percent. Obviously, it is a matter of interpretation whether this
low percentage is to be seen as ``slow'' star formation: especially at the lower density
thresholds, the freefall time $\tau_{ff}$ is
{\em not} representative of the local conditions under which the ``stars'' form. 
At later times, several effects are discernible. For low density thresholds (i.e. considering the ``whole''
cloud, see also Fig.~\ref{f:taufftime}), still $SFR_{ff}\lesssim 10$\%. For higher density thresholds
(i.e. selecting for the regions that are locally collapsing in the simulation), the $SFR_{ff}$ increases
but stays short of $\approx 50$\%. This is not surprising, 
since (a) these regions are not resolved, i.e. they would not convert all their mass into stars 
so that the ``real'' $SFR_{ff}$ would be substantially lower and 
(b) our models do not include feedback. Moreover, the models discussed here are an extreme realization
of a range of scenarios whose other extreme would be a perfect shear flow (Williams et al., in preparation). 
With a larger component of shear or angular momentum, 
the $SFR_{ff}$ could be reduced even further.

\subsection{Other limitations}
In addition to limited resolution, our models do not include stellar energy
input, which is the the most plausible mechanism to reverse the collapse of our
self-gravitating molecular clouds.  It is difficult to see how stellar energy input
can exactly balance the non-linear variation of gravity with position within the
cloud; there is more parameter space for either expansion or contraction.  
We note that even at the end of our simulations, most of the
mass of the cloud, including that of the molecular (CO) regions, has long free-fall
times and low densities.  The low filling factors indicated
by the $SFR_{ff}$ at $n\approx 100$~cm$^{-3}$ and by Figure~\ref{f:taufftime} suggest
that stellar feedback can in principle blow away most of the cloud mass and thus keep overall
star formation efficiencies low (see also \citealp{2000ApJ...530..277E}).  
While this remains an important issue of our scenario
which needs further investigation, it should also be pointed out that to our knowledge
there are no simulations of long-lived clouds that are {\em not} computed with periodic
boundary conditions, the adoption of which begs the question.

Magnetic fields could in principle suppress or affect the fragmentation 
(e.g. \citealp{1965ApJ...142..531F} for the thermal instability; \citealp{2007ApJ...665..445H}
for the dynamical instabilities in colliding flows), although
this capability seems to depend strongly on the number of dimensions considered 
(e.g. \citealp{2008arXiv0801.0486I}). Also, the sheer accumulation of mass along fieldlines
might render the fields less important than commonly thought, as demonstrated in 
numerical models by \citet{2008A&A...486L..43H}.

\section{Summary}\label{s:summary}

Making a simple but reasonable approximation for the formation of CO from atomic
gas, we have shown that the formation of molecular clouds from large-scale flows
can be tied closely to the formation of self-gravitating structures.  Our numerical
simulations support many parts of the rapid cloud and star formation scenario, 
in particular the rapid appearance of CO in self-gravitating clouds, quickly followed
by the formation of dense cores with free-fall timescales of order 1 Myr, and that
global gravitational collapse plays an important role in cloud evolution.
The simulations reproduce expected low star formation rates at early times. For
the long-term evolution of the star formation history and the molecular cloud
itself, the issue of stellar energy input (feedback) needs to be addressed to
develop a more comprehensive picture of star formation in the solar neighborhood.

\acknowledgements 
We thank the referee, Ian Bonnell, for the insightful and helpful report, and
Jim Pringle for illuminating discussions. The arguments presented lean heavily on
the evidence from numerical simulations performed at the National Center for
Supercomputing Applications (AST 060034) and on the local PC-cluster Star, perfectly
administered and maintained by J.~Hallum. 
FH is supported by the University of Michigan and NASA Herschel Cycle-0 grant NHSC1008.
This work has made use of NASA's Astrophysics Data System.

\bibliographystyle{apj}
\bibliography{./references}

%%%%%%%%%%%%%%%%%%%%%%%%%%%%%%%%%%%%%%%%%%%%%%%%%%
%%%%%%%%%%%%%%%%%%%%%%%%%%%%%%%%%%%%%%%%%%%%%%%%%%
\end{document}